\documentclass[a4paper]{amsart}

\usepackage[hypertexnames=false,linktocpage=true]{hyperref} 
\hypersetup{colorlinks=true,linkcolor=blue}

\usepackage{graphics,amssymb,enumerate}
\usepackage{hyperref}
\usepackage{graphicx}
\usepackage{float}
\floatstyle{boxed}
\restylefloat{figure}

\newcommand{\Ref}[1]{(\ref{#1})}
\newcommand{\tr}[1]{\operatorname{tr}(#1)}
\newcommand{\be}{\begin{equation}}
\newcommand{\ee}{\end{equation}}

\newtheorem{example}{Example}

\allowdisplaybreaks

\begin{document}
\title{Bi-Hamiltonian structures of Toda type systems}

\thanks{This work was co-funded by the European Regional Development Fund and the Re\-pu\-blic of Cyprus through the Research Promotion Foundation (Project: PENEK/0311/30).}

\author{C. A. Evripidou}
\address{Department of Mathematics and Statistics\\
University of Cyprus\\
P.O.~Box 20537, 1678 Nicosia\\Cyprus}
\email{evripidou.charalambos@ucy.ac.cy}

\date{}

\begin{abstract}
In this paper we define a family of systems which have similarities with the Toda lattice.
We construct two Lax pair representations and the associate Poisson structures for these systems.
These systems lie between the classical Toda lattice and the full Kostant-Toda lattice.
A Hamiltonian and a bi-Hamiltonian structure for these systems is constructed and also master symmetries for some low dimensional cases  are presented.
\end{abstract}

\maketitle

\section{Introduction} \label{section:introduction}

The Toda lattice is a completely integrable system introduced by Morikazu Toda, in \cite{toda}.
It was studied in several articles including \cite{damianou90}, \cite{damianou91}, \cite{damianou93}, \cite{damianou02}, \cite{flaschka2}, \cite{flaschka}, \cite{henon},  \cite{manakov}, \cite{moser74}. It is described by the equations

\be
\label{Toda-Lattice_symplectic}
\begin{array}{lcl}
\dot q_j=p_j, \,\, j=1,2,\ldots, n  \\
\dot p_j=e^{ q_{j-1}-q_j }- e^{q_j- q_{j+1}}, \,\, j=1,2,\ldots , n\ .
\end{array}
\ee
The system (\ref{Toda-Lattice_symplectic}) is Hamiltonian with Hamitonian function
$$
H_2=\frac{1}{2}\sum_{i=1}^n p_i^2 + \sum _{i=1}^{n-1}e^{ q_i-q_{i+1}}
$$
and bracket the symplectic bracket defined in $\mathbb{R}^{2n}$ by
\be
\{ f, g \}_s = \sum_{i=1}^n \left( {\partial f \over \partial q_i} {\partial g
\over \partial p_i} - {\partial f \over \partial p_i} { \partial g \over
\partial q_i} \right).
\label{Symplectic_bracket}
\ee

Under Flaschka's transformation
\begin{gather}
\begin{split}
a_i = {1 \over 2} e^{ {1 \over 2} (q_i - q_{i+1} ) }, \,\, i=1,2,\ldots, n-1\\
b_i =-{ 1 \over 2} p_i, \,\, i=1,2,\ldots,n,
\label{Flashka_transformation}
\end{split}
\end{gather}
the system (\ref{Toda-Lattice_symplectic}) transforms to
\be
\begin{array}{lcl}
\dot a _i& = & a_i \,  (b_{i+1} -b_i ), i=1,2,\ldots, n-1    \\
\dot b _i &= & 2 \, ( a_i^2 - a_{i-1}^2 ), i=1,2,\ldots, n,
\label{Toda_lattice}
\end{array}
\ee
where $a_0=0$.

Flaschka's transformation is useful for finding integrals of motion for the Toda lattice.
It can be easily verified that equations (\ref{Toda_lattice}) are
equivalent to the Lax equation $\dot{L}=[B,L]$ where
$$
L=
\begin{pmatrix}
b_1 & a_1 & 0 & \ldots & 0 & 0\\
a_1 & b_2 & a_2 &   & 0 & 0\\
0 & a_2 & b_3 & \ddots & \vdots & \vdots \\
\vdots & \vdots & \ddots & \ddots & \ddots & \vdots \\
0 & 0 & 0 & \ddots & b_{n-1} & a_{n-1} \\
0 & 0 & 0 & \cdots & a_{n-1} & b_n
\end{pmatrix}
$$
and $B$ is the skew-symmetric part of $L$
$$
B=
\begin{pmatrix}
0 & a_1 & 0 & \ldots & 0 & 0\\
-a_1 & 0 & a_2 &   & 0 & 0\\
0 & -a_2 & 0 & \ddots & \vdots & \vdots \\
\vdots & \vdots & \ddots & \ddots & \ddots & \vdots \\
0 & 0 & 0 & \ddots & 0 & a_{n-1} \\
0 & 0 & 0 & \cdots & -a_{n-1} & 0
\end{pmatrix}.
$$
The Jacobi matrix $L$ has the property that its eigenvalues are invariant over time
and therefore the functions $H_i=\tr{L^i}$ are constants of motion for the Toda lattice.

There exists a bracket in Flaschka's coordinates, which comes from the symplectic bracket and is defined by the relation
$$
\{f, g \} \circ F  = \frac{1}{4} \{ f \circ F, g \circ F \}_s \ .
$$
It is given by the equations
\begin{gather}
\begin{split}
\{a_i,b_i\} = -a_i, i=1,2,\ldots,n-1,\\
\{a_i,b_{i+1}\} = a_i, i=1,2,\ldots,n-1,
\label{linear_bracket_of Toda_lattice}
\end{split}
\end{gather}
and all other brackets are zero. We denote this bracket by $\pi_1$.
The function $H_1=b_1+b_2 + \dots +b_n$ is the only Casimir for this bracket and the Hamiltonian
$
H_2 = { 1 \over 2}\  { \rm tr}\  L^2=\frac{1}{2}(b_1^2+\ldots+b_n^2)+a_1^2+\ldots+a_{n-1}^2
$ gives equations (\ref{Toda_lattice}).
The functions $H_2,\ldots, H_n$ are enough to ensure the integrability of the system.
They are independent and in involution (i.e. $ \{  H_i, H_j \}=0$).

The Toda lattice is a bi-Hamiltonian system, meaning that there exist another Poisson bracket,
which we denote by $\pi_2$, and another function $F$, which will play the role of the Hamiltonian for the $\pi_2$ bracket, such that $\pi_1+\pi_2$ is Poisson and
$$
\pi_1\nabla H_2=\pi_2\nabla F.
$$
For the Toda lattice the $\pi_2$ bracket which appeared in a paper of Adler  \cite{adler79} in 1979
is quadradic in the variables $b_i, a_i$. It is given by the relations

\be
\begin{array}{lcl}
\{a_i, a_{i+1} \}&=&{ 1 \over 2} a_i a_{i+1} \\
\{a_i, b_i \} &=& -a_i b_i                    \\
\{a_i, b_{i+1} \}&=& a_i b_{i+1}    \\
\{b_i, b_{i+1} \}&=& 2\, a_i^2
\label{quadradic_Toda_bracket}
\end{array}
\ee
and all other brackets are zero. The Casimir for this bracket is the function ${\rm det}L$ and
$$
F=H_1 =\tr{L}=b_1+\ldots+b_n
$$
is the Hamiltonian and
$$
\pi_2 \nabla H_1 = \pi_1 \nabla H_2.
$$
Furthermore, $\pi_2$ is compatible with $\pi_1$. The functions $H_i$ are still in involution.

In this paper we will be concerned with a family of systems which are similar to the Toda lattice.
They are defined by lax equations of the form $\dot{L}=[B,L]$
where $L$ is similar to the Lax matrix for the Toda lattice
(see (\ref{lax_matrix_for_the_toda_type})).
This family of systems is also bi-Hamiltonian. It is closely related to a family of systems
which are called generalized Lotka-Volterra systems (see \cite{mine2015}). In section
\ref{section:generalized_lotka_volterra_systems} we recall the definitions and basic properties of
the generalized Lotka-Volterra systems. In section \ref{section:toda_type_systems}
we define this new family of Toda-type systems and we present a bi-Hamiltonian structure
for these systems. We also find master symmetries for some low dimensional examples.

\section{Generalized Lotka Volterra systems}
\label{section:generalized_lotka_volterra_systems}

The Toda lattice is closely related to the KM system.
The KM system is a well-known integrable system defined by the equations
\be
\label{a1}
\dot x_i = x_i(x_{i+1}-x_{i-1}) \qquad i=1,2, \dots,n,
\ee
where $x_0 = x_{n+1}=0$. It was studied by
Lotka in \cite{lotka} to model oscillating chemical reactions and by
Volterra in \cite{volterra} to describe population evolution in a
hierarchical system of competing species. It was first solved by
Kac and van-Moerbeke in \cite{kac}, using a discrete version of
inverse scattering due to Flaschka \cite{flaschka}. In
\cite{moser} Moser gave a solution of the system using the method
of continued fractions and in the process he constructed
action-angle coordinates. Equations \Ref{a1} can be considered
as a finite-dimensional approximation of the Korteweg-de Vries
(KdV) equation.  The Poisson bracket  for this system can
be thought as a lattice generalization of the Virasoro algebra
\cite{fadeev2}.

The Volterra system is associated with a simple Lie algebra of type $A_n$.
The KM system was generalized in \cite{mine2015}. For each subset of the positive roots of the root system of type $A_n$
there is associated a Hamiltonian system. For the simple roots of $A_n$ the system associated is the KM system.

The KM-system given by equation \Ref{a1} is Hamiltonian (see \cite{damianou91}, \cite{fadeev})
and can be written in Lax pair form in various ways. There is a symmetric version due to Moser where the matrices $L,B$ have the form
\begin{equation} \label{lax-symmetric}
L=
\begin{pmatrix}
0 &  a_1 & 0 & \cdots & \cdots & 0 \\
a_1 & 0 & a_2 & \ddots &    & \vdots \\
0 & a_2 & 0 & \ddots &  &  \vdots \\
\vdots & \ddots & \ddots & \ddots & & 0 \\
\vdots & & & \ddots & \ddots & a_n \\
0 & \cdots &  \cdots & 0 & a_{n} & 0
\end{pmatrix},
\end{equation}
and
\begin{equation*}
B=
\begin{pmatrix}
0 & 0  & a_1 a_2 & \cdots & \cdots & 0 \\
0 & 0 &  0 & \ddots &    & \vdots \\
-a_1 a_2 & 0 & 0 & \ddots & a_2 a_3 &  \vdots \\
\vdots & -a_2 a_3 & \ddots & \ddots & & a_{n-1} a_n \\
\vdots & & & \ddots & \ddots & 0 \\
0 & \cdots &  \cdots & -a_{n-1}a_n & 0 & 0
\end{pmatrix}.
\end{equation*}
\noindent
The matrix equation $\dot{L}=[B, L]$ gives a polynomial (in fact cubic) system of differential equations.
The change of variables $x_i=2a_i^2$ gives equations \Ref{a1}.

The previous Lax pair can be constructed using the following procedure (see \cite{damianou12}).
Let $\mathfrak{g}$ be any simple Lie algebra equipped with its Killing form $\langle\cdot\,\vert\,\cdot\rangle$. One chooses
a Cartan subalgebra $\mathfrak{h}$ of $\mathfrak{g}$, and a basis $\Pi$ of simple roots for the root system $\Delta$ of $\mathfrak{h}$ in
$\mathfrak{g}$. The corresponding set of positive roots is denoted by $\Delta^+$.  To each positive root $\alpha$ one can
associate a triple $(X_\alpha,X_{-\alpha},H_{\alpha})$ of vectors in $\mathfrak{g}$ which generate a Lie subalgebra
isomorphic to $sl_2(\mathbf{C})$. The set $(X_\alpha, X_{-\alpha})_{\alpha \in \Delta^+}\cup (H_\alpha)_{\alpha \in
\Pi}$ is a basis of $\mathfrak{g}$, called a root basis.
Let  $\Pi=\{ \alpha_1, \dots, \alpha_{\ell} \}$ and let    $X_{\alpha_1}, \ldots, X_{\alpha_\ell}$ be the corresponding root vectors in  $\mathfrak{g}$.  Define
\begin{displaymath} L=\sum_{\alpha_i \in \Pi} a_i (X_{\alpha_i}+X_{-\alpha_i})   \ . \end{displaymath}
To find the matrix $B$ we use the following procedure. For each $i,j$  form the vectors
$\left[X_{\alpha_i},X_{\alpha_j}\right]$.  If $\alpha_i+\alpha_j $ is a root then
 include a term of the form $a_i a_j \left[X_{\alpha_i},X_{\alpha_j}\right]$ in $B$.
We make $B$ skew-symmetric by including the corresponding negative  root vectors $a_i a_j [X_{-\alpha_i},X_{-\alpha_j}]$. Finally, we define the system using the Lax pair equation
\begin{displaymath} \dot{L}=[L, B]  \ . \end{displaymath}
For a root system of type $A_n$ we obtain the KM system.

In \cite{mine2015} the previous algorithm was generalized as follows.
Consider a subset $\Phi$ of the positive roots $\Delta^{+}$ of a root system of a simple Lie algebra such that
$$
\Pi \subset \Phi \subset \Delta^{+},
$$
where $\Pi$ is the set of simple roots.

The Lax matrix is defined as
\begin{displaymath}
L=\sum_{\alpha_i \in \Phi} a_i (X_{\alpha_i}+X_{-\alpha_i})   \ .
\end{displaymath}
Here we use the following enumeration of $\Phi$ which we assume to have $m$ elements.
The variables $a_j$ correspond to the simple roots $\alpha_j$ for $j=1,2, \dots, \ell$.
We assign the variables $a_j$  for $j=\ell+1, \ell+2, \dots, m $ to the remaining roots in $\Phi$.
The matrix $B$ is constructed using the following algorithm.
Consider the set $\Phi \cup \Phi^{-}$ which consists of all the roots in $\Phi$ together with their negatives.
Let
$$
\Psi =\left\{ \alpha+\beta \ | \ \alpha,  \beta \in \Phi \cup \Phi^{-}, \ \  \alpha+\beta \in \Delta^{+} \right\}
$$
and define
\be
\label{construction of B}
B=\sum c_{ij} a_i a_j (X_{\alpha_i+\alpha_j}-X_{-\alpha_i - \alpha_j}),
\ee
where $c_{ij}=\pm 1$  if  $\alpha_i+\alpha_j \in \Psi$ with $\alpha_i,\alpha_j\in\Phi\cup\Phi^-$ and $0$ otherwise.

\begin{example}
For each $n$ the algorithm constructed in \cite{mine2015} produce the KM system.
Let $E$ be the hyperplane  of $\mathbb{R}^{n+1}$ for which the coordinates sum to $0$ (i.e. vectors orthogonal to $(1,1,\ldots,1)$).
Let $\Delta$ be the set of vectors in $E$ of length $\sqrt{2}$ with integer coordinates. There are $n(n-1)$ such vectors in all.
We use the standard inner product in $\mathbb{R}^{n+1}$ and the standard orthonormal basis
$\{ \epsilon_1, \epsilon_2, \ldots,  \epsilon_{n+1} \}$.
Then, it is easy to see that $\Delta = \{ \epsilon_i-\epsilon_j \ | \ i \not= j \}$.
The vectors
\begin{displaymath}
\begin{array}{lcl}
\alpha_1 & =& \epsilon_1 -\epsilon_2 \\
\alpha_2 & =& \epsilon_2 -\epsilon_3 \\
 & \cdots &\\
\alpha_n & =& \epsilon_n -\epsilon_{n+1} \
\end{array}
\end{displaymath}
form a basis of the root system in the sense that each vector in $\Delta$ is a linear combination of these $n$ vectors with integer coefficients,
either all nonnegative or all nonpositive. For example,
$\epsilon_1 -\epsilon_3=\alpha_1+\alpha_2$, $ \epsilon_2 -\epsilon_4 =\alpha_2+\alpha_3$ and
$\epsilon_1-\epsilon_n=\alpha_1+\alpha_2+\ldots+\alpha_{n-1}$.
Therefore $\Pi=\{\alpha_1, \alpha_2, \ldots, \alpha_n \}$,
and the set of positive roots $\Delta^{+}$ is given by
\begin{displaymath}
\Delta^{+}= \{ \alpha_i+\ldots+\alpha_j:1\leq i\leq j\leq n  \} \ .
\end{displaymath}
If we take $\Phi=\Pi=\{ \alpha_1, \alpha_2,\ldots, \alpha_n \}$
then
$$
\Phi \cup \Phi^-=\{ \pm\alpha_1, \pm\alpha_2,\ldots, \pm\alpha_n \}
$$
and
$\Psi =\{ \alpha_i+\alpha_{i+1}: i=1,2,\ldots,n-1\}$.
We obtain the following Lax pair.
The matrix $L$ is
\label{ex1}
\begin{gather*}
L=
\sum_{\alpha_i \in \Phi} a_i (X_{\alpha_i}+X_{-\alpha_i})=
\sum_{i=1}^n a_i (X_{\alpha_i}+X_{-\alpha_i})=\\
\begin{pmatrix}
0 &  a_1 & 0 & \cdots & \cdots & 0 \\
a_1 & 0 & a_2 & \ddots &    & \vdots \\
0 & a_2 & 0 & \ddots &  &  \vdots \\
\vdots & \ddots & \ddots & \ddots & & 0 \\
\vdots & & & \ddots & \ddots & a_n \\
0 & \cdots &  \cdots & 0 & a_{n} & 0
\end{pmatrix}
\end{gather*}
which is the $L$ matrix of the Lax pair of the KM system.
Using the algorithm (\ref{construction of B}) (for $c_{1,2}=c_{2,3}=\ldots=c_{n-1,n}$) the matrix $B$ is
\begin{gather*}
B=
\begin{pmatrix}
0 & 0  & a_1 a_2 & \cdots & \cdots & 0 \\
0 & 0 &  0 & \ddots &    & \vdots \\
-a_1 a_2 & 0 & 0 & \ddots & a_2 a_3 &  \vdots \\
\vdots & -a_2 a_3 & \ddots & \ddots & & a_{n-1} a_n \\
\vdots & & & \ddots & \ddots & 0 \\
0 & \cdots &  \cdots & -a_{n-1}a_n & 0 & 0
\end{pmatrix}.
\end{gather*}
The Lax pair gives the system
\[
\begin{split}
\dot{a_1} & = a^2_2a_1,         \\
\dot{a_i} & =a_i(a^2_{i+1}-a^2_{i-1}), i=2,3,\ldots,n-1,\\
\dot{a_3} & =-a^2_{n-1}a_n.\\
\end{split}
\]
Using  the substitution $x_i=a_i^2$ followed by scaling  we obtain the KM system.
\[
\begin{split}
\dot{x_1} &=x_1x_2,\\
\dot{x_i}&=x_i(x_{i+1}-x_{i-1}), i=2,3,\ldots,n-1,\\
\dot{x_n}&=-x_{n-1}x_n.\\
\end{split}
\]
The KM system is integrable.
\end{example}

Moser in \cite{moser} describes a relation between the KM system \Ref{a1}
and the Toda lattice (\ref{Toda_lattice}). The procedure is the following.
Form $L^2$  which is not anymore a tridiagonal matrix but is similar to one.
Let $\{e_1, e_2, \dots, e_n \}$ be the standard basis of ${\bf R}^n$,
and $E_o=  {\rm span}\, \{  e_{2i-1}, \,  i=1,2, \dots \} ,
E_e=  {\rm span}\, \{ e_{2i}, \, i=1,2, \dots \}$. Then $L^2$ leaves $E_o$ and $E_e$ invariant
and reduces in each of these spaces to a tridiagonal symmetric Jacobi matrix.
For example, if we omit  all even columns and all even  rows we
obtain a tridiagonal Jacobi matrix and the entries of this new matrix
define the transformation from the KM--system to the Toda lattice.
We illustrate with a simple example where $n=5$.

We use the symmetric version of the KM system Lax pair given by
$$
L=\begin{pmatrix}
	0   & a_1 & 0   &  0  &  0  \\
	a_1 & 0   & a_2 &  0  &  0  \\
	0   & a_2 & 0   & a_3 &  0  \\
	0   & 0   & a_3 &  0  & a_4 \\
	0   & 0   & 0   & a_4 &  0
\end{pmatrix}  \ .
$$
It is simple to calculate that  $L^2$ is the matrix
$$
\begin{pmatrix}
a_1^2   &     0       & a_1 a_2     &     0       &    0    \\
0       & a_1^2+a_2^2 &     0       &   a_2 a_3   &    0    \\
a_1 a_2 &     0       & a_2^2+a_3^2 &      0      & a_3 a_4 \\
0       & a_2 a_3     &    0        & a_3^2+a_4^2 &    0    \\
0       &     0       & a_3 a_4     &      0      &   a_4^2
\end{pmatrix} \ .
$$
Omitting even  columns and even rows of $L^2$ we obtain the matrix
$$
\begin{pmatrix}
a_1^2   &    a_1 a_2  &    0    \\
a_1 a_2 & a_2^2+a_3^2 & a_3 a_4 \\
0       &   a_3 a_4   &  a_4^2
\end{pmatrix} \ .
$$
This is a tridiagonal Jacobi matrix. It is natural to define new variables
$A_1=a_1 a_2$, $A_2=a_3 a_4$, $B_1=a_1^2$, $B_2=a_2^2+a_3^2$, $B_3=a_4^2$.
The new variables $A_1,A_2, B_1,B_2, B_3$ satisfy the Toda lattice  equations.

This procedure shows that the KM-system  and the Toda lattice are closely related.
The explicit  transformation which is  due to H\'enon maps one system to the other.
The mapping in the general case  is given by
\be
A_i=-{ 1 \over 2} \sqrt {a_{2i} a_{2i-1}} \  ,
\qquad  B_i= { 1 \over 2}\left( a_{2i-1}+a_{2i-2} \right)
\label{a25} \ .
\ee
The equations satisfied by the new variables $A_i$, $B_i$ are given by
\begin{displaymath}
\begin{array}{lcl}
 \dot A _i& = & A_i \,  (B_{i+1} -B_i )    \\
   \dot B _i &= & 2 \, ( A_i^2 - A_{i-1}^2 ) \ .
\end{array}
\end{displaymath}
These are precisely the Toda equations \Ref{Toda_lattice} in Flaschka's form.

This idea of Moser was applied with success to establish transformations   from the generalized Volterra  lattices of Bogoyavlensky \cite{bog1, bog2}  to generalized Toda systems.
The relation between the Volterra systems of type $B_n$ and $C_n$ and the corresponding Toda systems is in \cite{damianou02}.  The similar construction of  the Volterra lattice   of type $D_n$ and the generalized
Toda lattice of type $D_n$   is in \cite{damianou04}.
The method of Moser was used in \cite{mine2015} to relate a family of generalized Lotka-Volterra systems
with a family of Toda type systems and
this relation was used to obtain a missing integral for that family of generalized Lotka-Volterra systems.

\section{Toda type systems}
\label{section:toda_type_systems}
For each $n\geq5$ we consider a system of $2n+2$ variables given by the equations

\begin{gather}
\begin{split}
\label{Toda_like_system}
\dot{b_1}&=2a_1^2-2a_n^2+2a_{n+2}^2,\\
\dot{b_2}&=2a_2^2-2a_1^2-2a_{n+1}^2,\\
\dot{b_i}&=2a_i^2-2a_{i-1}^2, \, i=3,\ldots,n-1,\\
\dot{b_{n-1}}&=2a_{n-1}^2-2a_{n-2}^2+2a_{n+1}^2,\\
\dot{b_n}&=2a_{n+1}^2-2a_{n-1}^2-2a_{n+2}^2,\\
\dot{a_i}&=a_i(b_{i+1}-b_i), \, i=1,\ldots,n-1,\\
\dot{a_n}&=a_n(b_{1}-b_{n-1})+2a_{n-1}a_{n+2},\\
\dot{a_{n+1}}&=a_{n+1}(b_{2}-b_{n})-2a_{1}a_{n+2},\\
\dot{a_{n+2}}&=a_{n+2}(b_{1}-b_{n})+2a_{1}a_{n+1}-2a_{n-1}a_n.
\end{split}
\end{gather}

These systems are Hamiltonian with Hamiltonian function
$$
H=b_1+\ldots+b_n
$$
and Poisson bracket defined by the relations
\begin{gather*}
\{b_i,b_{i+1}\}=2a_i^2, \, i=1,\ldots,n-1,\\
\{b_1,b_n\}=2a_{n+2}^2,  \,\,\, \{b_i,b_{n-2+i}\}=-2a_{n-1+i}^2, \, i=1,2,\\
\{b_i,a_j\}=b_ia_j, \text{ for } (i,j)=(1,2), \ldots, (n-1,n), (1,n+2), (n-1,n), (n,n+1),\\
\{b_i,a_j\}=-b_ia_j, \text{ for } (i,j)=(2,1), \ldots, (n,n-1), (1,n), (2,n+1), (n,n+2),\\
\{b_1,a_{n+1}\}=2a_1a_{n+2}, \{b_{n-1},a_{n+2}\}=2a_{n+1}a_{n+2},\\
\{b_2,a_{n+2}\}=-2a_1a_{n+1}, \{b_{n},a_{n}\}=-2a_{n-1}a_{n+2},\\
\{a_i,a_{j}\}=\frac{1}{2}a_ia_j \text{ for } (i,j)=(1,2), \ldots, (n-2,n-1),\\
(1,n+1), (n-1,n+1), (n-1,n+2), (n-2,n)\\
\{a_i,a_{j}\}=-\frac{1}{2}a_ia_j \text{ for } (i,j)= (1,n), (1,n+2), (2,n+1), (n-1,n),\\
\{a_n,a_{n+1}\}=a_1a_{n-1}, \{a_n,a_{n+2}\}=\frac{1}{2} a_na_{n+2}+b_1a_{n-1},\\ 
\{a_{n+1},a_{n+2}\}=-\frac{1}{2}a_{n+1}a_{n+2}-a_1b_n.
\end{gather*}
Note that setting $a_n=a_{n+1}=0$ we get the equations of the periodic Toda lattice, while if we set $a_n=a_{n+1}=a_{n+2}=0$
we get the equations (\ref{Toda_like_system}).

The system (\ref{Toda_like_system}) can be written in Lax pair form $\dot{L}=[B,L]$ with

\begin{gather}
L=
\begin{pmatrix}
b_1 & a_1 & 0 & \ldots & a_n & a_{n+2}\\
a_1 & b_2 & a_2 &   & 0 & a_{n+1}\\
0 & a_2 & b_3 & \ddots & \vdots & 0 \\
\vdots & \ddots & \ddots & \ddots & \ddots & \vdots \\
a_n & 0 & 0 & \ddots & b_{n-1} & a_{n-1} \\
a_{n+2} & a_{n+1} & 0 & \cdots & a_{n-1} & b_n
\end{pmatrix}
\label{lax_matrix_for_the_toda_type}
\end{gather}
and
$$
B=
\begin{pmatrix}
0 & a_1 & 0 & \ldots & -a_n & a_{n+2}\\
-a_1 & 0 & a_2 &   & 0 & -a_{n+1}\\
0 & -a_2 & 0 & \ddots & \vdots & 0 \\
\vdots & \ddots & \ddots & \ddots & \ddots & \vdots \\
a_n & 0 & 0 & \ddots & 0 & a_{n-1} \\
-a_{n+2} & a_{n+1} & 0 & \cdots & -a_{n-1} & 0
\end{pmatrix}
$$

The system (\ref{Toda_like_system}) can be obtained from a generalized Lotka-Volterra system produced in \cite{mine2015}
using the procedure of Moser. The generalized Lotka-Volterra system is defined by the Lax equation $\dot{L}=[B,L]$
where the matrix $L$ is defined by

$$
L=
\begin{pmatrix}
0     & a_1   & 0    &\cdots&  0    &  a_n  & 0       \\
a_1   & 0     & a_2  &\ddots&       &  0    &a_{n+1}  \\
0     & a_2   &  0   &\ddots&       &       &0        \\
\vdots&\ddots &\ddots&\ddots&       &   0   &\vdots   \\
 0    &       &      &      &       &a_{n-2}&0        \\
a_n   &  0    &      &  0   &a_{n-2}&  0    &a_{n-1}  \\
0     &a_{n+1}&  0   &\cdots&   0   &a_{n-1}&0
\end{pmatrix}
$$
and corresponds to the subset $\Phi$ of the positive roots containing the simple roots and the roots of length $n-2$.
The matrix $B$ is defined by equation (\ref{construction of B}) and its upper triangular part is
\arraycolsep=1pt\def\arraystretch{1}
$$
\begin{pmatrix}
0      &0      &a_1a_2 &0              &\cdots&   0     & -a_{n-2}a_{n} &0     &a_1a_{n+1}+a_{n-1}a_n \\
0      &0      &0      &a_2a_3         &      &   \;    &      0        &-a_1a_{n}-a_{n-1}a_{n+1}&0   \\
\vdots & \ddots&0      &0              &\ddots&   \;    &               &    0                   &-a_2a_{n+1}\\
       &       &       &\ddots         &\ddots&   \;    &               &                        &     0     \\
       &       &       &               &      &   \;    &   \ddots      &    0                   &    \vdots \\
       &       &       &               &      &   \;    &     0         &a_{n-3}a_{n-2}          &     0     \\
       &       &       &               &      &   \;    &     0         &0                       & a_{n-2}a_{n-1}\\
\vdots &       &       &               &      &         &   \ddots      & 0                      &0              \\
0      &\cdots &       &               &      &         &   \cdots      & 0                      &0
\end{pmatrix},
$$
\arraycolsep=4pt\def\arraystretch{1}
The Lax equation $\dot{L} =[B,L]$ is equivalent to the following system:
\begin{align}
\label{2-diagonal_system}
\begin{split}
	\dot{a}_1     &=	a_1a^2_2+a_1a^2_{n+1}-a_1a^2_n, \\
	\dot{a}_2     &=	a_2a^2_3-a^2_1a_2-a_2a^2_{n+1}, \\
	\vdots  \     & 	\quad\quad\quad\quad	\vdots  \\
	\dot{a}_i     &=	a_ia^2_{i+1}-a^2_{i-1}a_i, \quad \quad \quad i = 3, 4, \dots, n-3\\
	\vdots \      & 	\quad\quad\quad\quad	\vdots  \\
	\dot{a}_{n-2} &=	a_{n-2}a^2_n-a^2_{n-3}a_{n-2}+a_{n-2}a^2_{n-1},\\
	\dot{a}_{n-1} &=	a_{n-1}a^2_{n+1}-a^2_{n-2}a_{n-1}-a_{n-1}a^2_n,\\
	\dot{a}_n     &=	a^2_1a_n+a^2_{n-1}a_n-a^2_{n-2}a_n+2a_{1}a_{n-1}a_{n+1},\\
	\dot{a}_{n+1} &=a^2_2a_{n+1}-a^2_1a_{n+1}-a_{n+1}a^2_{n-1}-2a_1a_{n-1}a_n\,.
\end{split}
\end{align}

\noindent
The upper triangular part of the Poisson matrix of this system is
\be
\label{Poisson_matrix_of_2-diagonal_system}
\begin{pmatrix}
0      &a_1a_2 &    0  & \cdots        &      &       & 0             & -a_1a_n      & a_1a_{n+1}     \\
\vdots &0      &a_2a_3 &     0         &      & \;    &               & 0            & -a_2a_{n+1}    \\
       &       &0      & a_3a_4        &      & \;    &               &              &      0         \\
       &       &       & \ddots        &      & \ddots& 0             &              &                \\
       &       &       &               &\ddots& \ddots& 0             & 0            &    \vdots      \\
       &       &       &               &      & 0     & a_{n-2}a_{n-1}& a_{n-2}a_{n} &     0          \\
       &       &       &               &      & \;    & 0             & -a_{n-1}a_n  & a_{n-1}a_{n+1} \\
\vdots &       &       &               &      & \;    &               & 0            & 2a_1a_{n-1}    \\
0      &\cdots &       &               &      & \;    &               & \cdots       & 0
\end{pmatrix},
\ee

Using the procedure of Moser we define the matrix $L^2$ in which we omit its even rows and even columns to obtain the matrix

\begin{gather*}
\begin{pmatrix}
a_1^2+a_n^2 & a_1a_2 & 0 & \ldots & a_{n-2}a_n & a_1a_{n+1}+a_{n-1}a_n\\
a_1a_2 & a_2^2+a_3^2 & a_3a_4 &   & 0 & a_2a_{n+1}\\
0 & a_3a_4 & a_4^2+a_5^2 & \ddots & \vdots & 0 \\
\vdots & \ddots & \ddots & \ddots & \ddots & \vdots \\
a_{n-2}a_n & 0 & 0 & \ddots & a_{n-2}^2+a_{n}^2 & a_{n-2}a_{n-1} \\
a_1a_{n+1}+a_{n-1}a_n & a_2a_{n+1} & 0 & \cdots & a_{n-2}a_{n-1} & a_{n-1}^2+a_{n+1}^2
\end{pmatrix}\\
:=\begin{pmatrix}
B_1 & A_1 & 0 & \ldots & A_n & A_{n+2}\\
A_1 & B_2 & A_2 &   & 0 & A_{n+1}\\
0 & A_2 & B_3 & \ddots & \vdots & 0 \\
\vdots & \ddots & \ddots & \ddots & \ddots & \vdots \\
A_n & 0 & 0 & \ddots & B_{n-1} & A_{n-1} \\
A_{n+2} & A_{n+1} & 0 & \cdots & A_{n-1} & B_n
\end{pmatrix}
\end{gather*}
It is straightforward to verify that the new variables $B_1,\ldots,B_n,A_1,\ldots,A_{n+2}$
satisfy the equations (\ref{Toda_like_system}).

The system (\ref{2-diagonal_system}) is Hamiltonian with Hamiltonian function
$\sum_{i=1}^na_i^2$ and Poisson matrix (\ref{Poisson_matrix_of_2-diagonal_system}).
Using the Hamiltonian formulation of the system (\ref{2-diagonal_system})
we are able to define a bi-Hamiltonian formulation of the system (\ref{Toda_like_system}).

We define a second Poisson bracket $\pi_2$ for the system (\ref{Toda_like_system}) which is linear in the variables $a_i$.
\be
\label{Linear_Poisson_for_the_Toda_like}
\pi_2
=
\begin{pmatrix}
A & B \\
-B & C
\end{pmatrix}
\ee
where
$$
A=
\begin{pmatrix}
0&0&\ldots&0\\
0&0&\ldots&0\\
\vdots&\vdots&\ddots&\vdots\\
0&0&\ldots&0\\
\end{pmatrix}\in\mathbb{R}^{n\times n}
$$
$$
B=
\begin{pmatrix}
a_1&0&0&\cdots&\cdots&0&-a_n&0&a_{n+2}\\
-a_1&a_2&0&\cdots&\cdots&0&0&-a_{n+1}&0\\
0&-a_2&a_3&\cdots&\cdots&0&0&0&0\\
\vdots&\vdots&\ddots&\ddots&\cdots&\cdots&\vdots&\vdots&\vdots\\
0&0&0&\ddots&\ddots&\cdots&0&0&0\\
0&0&0&\cdots&-a_{n-2}&a_{n-1}&a_n&0&0\\
0&0&0&\cdots&0&-a_{n-1}&0&a_{n+1}&-a_{n+2}\\
\end{pmatrix}\in\mathbb{R}^{n\times{(n+2)}}
$$
and
$$
C=
\begin{pmatrix}
0&0&0&0&0&0&0\\
0&0&0&0&0&0&0\\
0&0&0&0&0&0&0\\
0&0&0&0&0&0&0\\
0&0&0&0&0&0&a_{n-1}\\
0&0&0&0&0&0&-a_1\\
0&0&0&0&-a_{n-1}&a_1&0
\end{pmatrix}
\in\mathbb{R}^{(n+2)\times(n+2)}
$$

Note that the matrix $A$ determines the brackets between the variables $b_i$, and therefore $\{b_i,b_j\}=0 \, \, \forall i,j$,
the matrix $B$ determines the brackets between the variables $b_i,a_j$
and the matrix $C$ determines the brackets between the variables $a_i$.

The Hamiltonian for this bracket is the function
$$
H_2=\frac{1}{2}(b_1^2+\ldots+b_n^2)+a_1^2+\ldots+a_{n+2}^2.
$$
The rank of the Poisson matrix (\ref{Linear_Poisson_for_the_Toda_like}) is $2n$ and therefore there are two Casimirs for this bracket.
The functions $H_1=\sum_{i=1}^nb_i$ and $C=\prod_{i=2}^{n-2} a_i(a_1a_n+a_{n-1}a_{n+1})$ are the Casimirs.
The functions $H_i=\tr{L^i}$ are constants of motion for the system
but not enough to ensure its integrability.
\begin{example}
For $n=5$ the equations (\ref{Toda_like_system}) are

\begin{align}
\begin{split}
\label{Equations_for_n=5}
\dot{b_1 } &=2a_1^2-2a_5^2+2a_7^2\\ 
\dot{b_2 } &=-2a_1^2+2a_2^2-2a_6^2\\ 
\dot{b_3 } &=-2a_2^2+2a_3^2\\ 
\dot{b_4 } &=2a_5^2-2a_3^2+2a_4^2\\ 
\dot{b_5 } &=-2a_7^2+2a_6^2-2a_4^2\\ 
\dot{a_1 } &=b_2a_1-b_1a_1\\ 
\dot{a_2 } &=b_3a_2-b_2a_2\\ 
\dot{a_3 } &=b_4a_3-b_3a_3\\ 
\dot{a_4 } &=b_5a_4-b_4a_4\\ 
\dot{a_5 } &=b_1a_5+2a_7a_4-b_4a_5\\ 
\dot{a_6 } &=-2a_1a_7+b_2a_6-b_5a_6\\ 
\dot{a_7 } &=-b_1a_7+2a_1a_6-2a_5a_4+b_5a_7.
\end{split}
\end{align}

This system is equivalent to the Lax equation $\dot{L}=[B,L]$ where

$$
L=
\begin{pmatrix}
b_1&a_1&0&a_5&a_7\\
a_1&b_2&a_2&0&a_6\\
0&a_2&b_3&a_3&0\\
a_5&0&a_3&b_4&a_4\\
a_7&a_6&0&a_4&b_5
\end{pmatrix}
$$
and
$$
B=
\begin{pmatrix}
0&a_1&0&-a_5&a_7\\
-a_1&0&a_2&0&-a_6\\
0&-a_2&0&a_3&0\\
a_5&0&-a_3&0&a_4\\
-a_7&a_6&0&-a_4&0
\end{pmatrix}.
$$
A bi-Hamiltonian formulation for this system is given by the Poisson matrices
\begin{equation}
\label{Quadradic_Poisson_matrix_for_n=5}
\pi_1=
\begin{pmatrix}
A_1&B_1\\
-B_1&C_1\\
\end{pmatrix}
\end{equation}
and
\begin{equation}
\label{Linear_Poisson_matrix_for_n=5}
\pi_2=
\begin{pmatrix}
A_2&B_2\\
-B_2&C_2\\
\end{pmatrix}
\end{equation}
where
$$
A_1=
\begin{pmatrix}
0&2a_1^2&0&-2a_5^2&2a_7^2\\ 
-2a_1^2&0&2a_2^2&0&-2a_6^2\\ 
0&-2a_2^2&0&2a_3^2&0\\ 
2a_5^2&0&-2a_3^2&0&2a_4^2\\ 
-2a_7^2&2a_6^2&0&-2a_4^2&0
\end{pmatrix},
$$
$$
B_1=
\begin{pmatrix}
b_1a_1&0&0&0&-b_1a_5&2a_1a_7&b_1a_7\\ 
-b_2a_1&b_2a_2&0&0&0&-b_2a_6&-2a_1a_6\\ 
0&-b_3a_2&b_3a_3&0&0&0&0\\ 
0&0&-b_4a_3&b_4a_4&b_4a_5&0&2a_5a_4\\ 
0&0&0&-b_5a_4&-2a_7a_4&b_5a_6&-b_5a_7
\end{pmatrix}
$$
\arraycolsep=1pt\def\arraystretch{1}
$$
C_1=-\frac{1}{2}
\begin{pmatrix}
0&-a_1a_2&0&0&a_1a_5&-a_1a_6&a_1a_7\\ 
a_1a_2&0&-a_2a_3&0&0&a_2a_6&0\\ 
0&a_2a_3&0&-a_3a_4&-a_5a_3&0&0\\ 
0&0&a_3a_4&0&a_5a_4&-a_6a_4&-a_7a_4\\ 
-a_1a_5&0&a_5a_3&-a_5a_4&0&-a_1a_4&-a_5a_7-a_4b_1\\ 
a_1a_6&-a_2a_6&0&a_6a_4&a_1a_4&0&a_7a_6+a_1b_5\\ 
-a_1a_7&0&0&a_7a_4&a_5a_7+a_4b_1&-a_7a_6-a_1b_5&0
\end{pmatrix}
$$
and
\arraycolsep=4pt\def\arraystretch{1}
$$
A_2=0
$$
$$
B_2=
\begin{pmatrix}
-a_1&0&0&0&a_5&0&-a_7\\ 
a_1&-a_2&0&0&0&a_6&0\\ 
0&a_2&-a_3&0&0&0&0\\ 
0&0&a_3&-a_4&-a_5&0&0\\ 
0&0&0&a_4&0&-a_6&a_7
\end{pmatrix}
$$
$$
C_2=
\begin{pmatrix}
0&0&0&0&0&0&0\\ 
0&0&0&0&0&0&0\\ 
0&0&0&0&0&0&0\\ 
0&0&0&0&0&0&0\\ 
0&0&0&0&0&0&-a_4\\ 
0&0&0&0&0&0&a_1\\ 
0&0&0&0&a_4&-a_1&0
\end{pmatrix}
$$
Note that the matrices $A_1,A_2$ determine the brackets between the variables $b_i$,
the matrices $B_1,B_2$ determine the brackets between the variables $b_i,a_j$
and the matrices $C_1,C_2$ determine the brackets between the variables $a_i$.

The Hamiltonian function for the bracket $\pi_1$ is
$$
H_1=b_1+\ldots+b_5
$$
and the determinant of the matrix $L$ is a Casimir.
The functions
$$
H_i=\tr	 L^i, \, i=2,3,4,5,6
$$
are constants of motion for the system (\ref{Equations_for_n=5}).
The Poisson matrix (\ref{Quadradic_Poisson_matrix_for_n=5}) has rank $10$ and therefore there is another Casimir for this bracket. 

The Hamiltonian function for the bracket $\pi_2$ is
$$
H_2=\frac{1}{2}(b_1^2+\ldots+b_5^2)+a_1^2+\ldots+a_7^2.
$$
The rank of the Poisson matrix (\ref{Linear_Poisson_matrix_for_n=5}) is $10$ and the functions
$H_1$ and $C=a_1a_2a_3a_5+a_2a_3a_4a_6$ are the Casimirs. The functions $H_i$ are constants of motion for the system
but not enough to ensure its integrability.

We can easily verify that the vector field 
$$
\mathcal{X}=\sum_{i=1}^5t_i\frac{\partial}{\partial b_i}+\sum_{i=1}^7t_{i+5}\frac{\partial}{\partial a_i}
$$
where
\begin{gather*}
t_1=\frac{1}{a_2a_6}(a_1^2a_6a_2+2a_1^2a_3a_4-2a_1a_3a_5a_6+a_5^2a_6a_2+a_7^2a_6a_2+b_1^2a_6a_2)\\
t_2=\frac{1}{a_2a_6}(a_1^2a_6a_2-2a_1^2a_3a_4+2a_1a_3a_5a_6+3a_2^3a_6+2a_2^2a_3a_4+a_6^3a_2+b_2^2a_6a_2)\\
t_3=-\frac{1}{a_6}(a_2^2a_6+2a_2a_3a_4-a_3^2a_6-b_3^2a_6)\\
t_4=(a_3^2+a_4^2+a_5^2+b_4^2)\\
t_5=(a_4^2+a_6^2+a_7^2+b_5^2)\\
t_6=\frac{1}{a_2a_6 \left(b_2-b_4\right)}(a_1a_2a_5^2a_6+b_1a_1a_6a_2b_2-\\
b_1a_1a_6a_2b_4+a_1b_2^2a_6a_2-a_1b_2a_6a_2b_4-a_1a_3a_4a_5^2-a_1a_3a_4b_1b_2+\\
a_1a_3a_4b_1b_4+a_1a_3a_4b_2^2-a_1a_3a_4b_2b_4+2a_2^2a_3a_5a_6+a_2a_3^2a_4a_5+\\
a_2a_4a_5a_6^2+a_7a_6^2a_2b_2-a_7a_6^2a_2b_4+a_3a_5^3a_6+a_3a_5a_6b_1b_2-\\
a_3a_5a_6b_1b_4-a_3a_5a_6b_2^2+a_3a_5a_6b_2b_4)\\
t_7=\frac{1}{a_2a_6\left(b_2-b_4\right)}(a_1a_2a_3a_5a_6-a_1a_3^2a_4a_5+\\
2a_2^2a_3^2a_6+2a_2^2b_3a_6b_2-2a_2^2b_3a_6b_4+a_2a_3^3a_4+a_2a_3a_4a_6^2-\\
a_2a_3a_4b_2^2+a_2a_3a_4b_2b_3+a_2a_3a_4b_2b_4-a_2a_3a_4b_3b_4+a_3^2a_5^2a_6)\\
t_8=-\frac{1}{a_6\left(b_2-b_4\right)}(a_1a_2a_5a_6-a_1a_3a_4a_5+2a_2^2a_3a_6+\\
a_2a_3^2a_4+a_2a_4a_6^2+a_3a_5^2a_6-b_3a_3a_6b_2-a_3b_4a_6b_2+b_3a_3a_6b_4+\\
a_3b_4^2a_6)\\
t_9=-\frac{1}{a_2\left(b_2-b_4\right)}(a_1a_2a_5a_6-a_1a_3a_4a_5+2a_2^2a_3a_6+a_2a_3^2a_4+\\
a_2a_4a_6^2-b_4a_4a_2b_2-a_4b_5a_2b_2+b_4^2a_4a_2+a_4b_5a_2b_4-a_5a_7a_2b_2+a_5a_7a_2b_4+a_3a_5^2a_6)\\
t_{10}=-\frac{1}{a_2a_6\left(b_2-b_4\right)}(a_1^2a_2a_5a_6-a_1^2a_3a_4a_5+2a_1a_2^2a_3a_6+\\
a_1a_2a_3^2a_4+a_1a_2a_4a_6^2+a_1a_3a_5^2a_6-a_7a_4a_6a_2b_2+a_7a_4a_6a_2b_4-b_1a_5a_6a_2b_2+\\
b_1a_5a_6a_2b_4-a_5b_4a_6a_2b_2+a_5b_4^2a_6a_2)\\
t_{11}=\frac{1}{a_2a_6\left(b_2-b_4\right)}(a_1a_2a_4a_5a_6+a_1a_7a_6a_2b_2-a_1a_7a_6a_2b_4-\\
a_1a_3a_4^2a_5-a_1a_3a_4a_7b_2+a_1a_3a_4a_7b_4+2a_2^2a_3a_4a_6+a_2a_3^2a_4^2+a_2a_4^2a_6^2+\\
b_2^2a_6^2a_2-b_2a_6^2a_2b_4+a_6^2b_5a_2b_2-a_6^2b_5a_2b_4+a_3a_4a_5^2a_6+a_3a_5a_6a_7b_2-a_3a_5a_6a_7b_4)\\
t_{12}=\frac{1}{a_2}(a_1a_6a_2+a_1a_3a_4+a_5a_4a_2+b_1a_7a_2+a_7b_5a_2-a_5a_3a_6)
\end{gather*}
is a master symmetry for this system which $\mathcal{X}(H_i)=H_{i+1}$.
\end{example}

\def\polhk#1{\setbox0=\hbox{#1}{\ooalign{\hidewidth
  \lower1.5ex\hbox{`}\hidewidth\crcr\unhbox0}}}

\end{document}